\documentclass[aps,pra,twocolumn,superscriptaddress,nofootinbib]{revtex4-1}
\usepackage{amsmath}
\usepackage{amssymb}
\usepackage{graphicx}
\usepackage{mathrsfs}
\usepackage{ntheorem}
\usepackage{mathtools}
\usepackage{enumerate}
\usepackage{enumitem}
\usepackage{times,txfonts}
\usepackage{subfigure}
\usepackage{upgreek}
\usepackage{bm}
\usepackage{tikz}
\usepackage[colorlinks,bookmarksopen,bookmarksnumbered,citecolor=blue, linkcolor=blue, urlcolor=blue]{hyperref}
\newcommand{\ket}[1]{|#1\rangle}
\newcommand{\bra}[1]{\langle #1|}
\newcommand{\Tr}{\mathrm{Tr}}

\newcommand{\abs}[1]{\lvert #1\rvert}

\def\CC{{\rm\kern.24em \vrule width.04em height1.46ex depth-.07ex \kern-.30em C}}
\def\RR{{\rm\kern.24em \vrule width.04em height1.46ex depth-.07ex
\kern-.30em R}}
\def\P{{\rm I\kern-.25em P}}

\begin{document}

\title{Approximate distillation of quantum coherence}

\author{C. L. Liu}
\affiliation{Graduate School of China Academy of Engineering Physics, Beijing 100193, China}

\author{C. P. Sun}
\email{suncp@gscaep.ac.cn}
\affiliation{Graduate School of China Academy of Engineering Physics, Beijing 100193, China}
\affiliation{Beijing Computational Science Research Center, Beijing 100193, China}
\date{\today}

\begin{abstract}
Coherence distillation is a basic information-theoretic task in the resource theory of coherence. In this paper, we develop the framework of the approximate coherence distillation under strictly incoherent operations. This protocol considers the situation that we cannot transform an initial state $\rho$ into a target state $\psi$ with certainty; instead, we aim to deterministically transform the initial state $\rho$ into an intermediate state $\phi$ that is most approximate to the target state $\psi$ in terms of fidelity. We also present the explicit conversion strategy of the approximate coherence distillation.
\end{abstract}
\maketitle

\section{Introduction}
Quantum coherence is a fundamental feature of quantum physics which is responsible for the departure between the classical and the quantum worlds. It is an essential component in quantum information processing \cite{Nielsen} and plays a central role in various fields, such as quantum computation \cite{Shor,Grover}, quantum cryptography \cite{Bennett}, quantum metrology \cite{Giovannetti,Giovannetti1}, and quantum biology \cite{Lambert}. Recently, the resource theory of coherence has attracted a growing interest due to the rapid development of quantum information science \cite{Aberg1,Baumgratz,Streltsov}. The resource theory of coherence establishes a rigorous framework to quantify coherence and provides a platform to understand  quantum coherence from a different perspective.

Any quantum resource theory is characterized by two fundamental ingredients: The free states and the free operations \cite{Chitambar}. For the resource theory of coherence, the free states are quantum states which are diagonal in a prefixed reference basis. The free operations are not uniquely specified. Motivated by suitable practical considerations, several free operations were presented \cite{Streltsov}, such as the maximally incoherent operations  \cite{Aberg1}, the incoherent operations  \cite{Baumgratz}, and the strictly incoherent operations  \cite{Winter,Yadin}. In this paper, we focus our attention on the strictly incoherent operations, which were first proposed in Ref. \cite{Winter} and it has been shown that these operations neither create nor use coherence and have a physical interpretation in terms of interferometry in Ref. \cite{Yadin}. Thus the set of strictly incoherent operations is a physically well-motivated set of free operations for the resource theory of coherence.

When we are performing a quantum information processing task, it is usually the pure coherent states that play the central role. Thus the ability of a quantum state to transform it into some pure coherent state is an important characterization of the given state. Since coherence distillation is the measure characterizing this ability, much effort has been devoted to investigate the distillation of coherence \cite{Chitambar}. Previous coherence distillation protocols can be divided into two different settings: The asymptotic regime
\cite{Yuan,Winter,Lami,Lami1,Zhao} and the one-shot regime \cite{Liu1,Liu2,Liu3,Fang,Regula,Regula1,Chitambar2,Torun,Regula2,Chen,Zhang,Du,Ziwen,Zhu,Du1,Liu4,Chitambar3}. For the coherence distillation protocols under strictly incoherent operations, the necessary and sufficient conditions for the asymptotic distillability were presented and the optimal rate for this distillation process was evaluated analytically in Refs. \cite{Lami,Lami1,Zhao}. In this case, the protocol has been assumed that there is an unbounded number of independent and identically distributed copies of a quantum state. The exact coherence distillations including the deterministic coherence distillation and the probabilistic coherence distillation  were studied in Refs. \cite{Liu1,Du,Zhu,Chitambar2,Liu2,Du1,Zhu,Chitambar2,Liu4,Torun}. In this exact case, we focus on the transformation that converts a collection of coherent states having different amounts of coherence into a target pure coherent state with certainty.

However, previous results about the exact coherence distillations show that the restriction to exact coherence distillation is too stringent. In other words, for an initial state $\rho$ and a target state $\psi$, the transformation from $\rho$ into $\psi$ may not be achievable by using strictly incoherent operations. In this case, we consider the approximation coherence distillation. For the approximate coherence distillation, instead of obtaining the target state $\psi$ exactly, we only need to get an intermediate state which has a high fidelity with the target state $\psi$. Here since it is usually the pure states that play the central role in quantum information processing tasks, we require the intermediate state to be a pure state or a pure state ensemble. In this protocol, the most fundamental questions are: For an initial state $\rho$ and the target state $\psi$, what is the maximal fidelity achievable in the approximate coherence distillation and what is the intermediate state achieving this maximal fidelity?

In this paper, we address the above questions by developing a general framework of approximate coherence distillation. More precisely, for an initial state $\rho$ and the target state $\psi$, we obtain the maximal fidelity achieved in the approximate coherence distillation and we also present the intermediate state that achieves the maximal fidelity. This distillation protocol can be seen as a generalization of the exact coherence distillation.

%The paper is organized as follows. In Sec.~II, we recall some notions of the resource theory of coherence and the notions of majorization and fidelity. In Sec.~III, we present the main result of this paper, i.e., the approximate coherence distillation. Section IV  offers our remarks and conclusions.

\section{Preliminaries}\label{II}

To present our result clearly, it is instructive to introduce some elementary notions of the resource theory of coherence \cite{Baumgratz}. Let $\{\ket{i}\}_{i=1}^d$ be the prefixed basis in the finite dimensional Hilbert space. A state is said to be incoherent if it is diagonal in the basis and the set of such states is denoted by $\mathcal{I}$. Coherent states are these not of this form. For a pure state $\ket{\varphi}$, we will write $\varphi:=\ket{\varphi}\bra{\varphi}$.

A strictly incoherent operation \cite{Winter, Yadin} is a completely positive trace preserving (CPTP) map, expressed as
\begin{eqnarray}
\Lambda(\rho)=\sum_{\mu=1}^N K_\mu\rho K_\mu^\dagger,
\end{eqnarray}
where the Kraus operators $K_\mu$ satisfy not only $\sum_{\mu=1}^N K_\mu^\dagger K_\mu=\mathbb{I}$ but also $K_\mu\mathcal{I}K_\mu^\dagger\subset \mathcal{I}~\text{and}~K_\mu^\dag\mathcal{I}K_\mu\subset \mathcal{I}$ for every $K_\mu$ \cite{Winter, Yadin}. One sees by inspection that there is at most one nonzero element in each column and row of $K_\mu$, and $K_\mu$ are called strictly incoherent operators. From this, it is elementary to show that a projector is incoherent if it is of the form $\mathbb{P}_\mu=\sum_{i}\ket{i}\bra{i}$ and we will denote $\mathbb{P}_\mu$ as a generic strictly incoherent projector. Hereafter, we will use $\Delta\rho=\sum_{i=1}^d\ket{i}\bra{i}\rho\ket{i}\bra{i}$ to denote the fully dephasing channel. A CPTP map is an incoherent operation if each $K_\mu$ only satisfies $K_\mu^\dag\mathcal{I}K_\mu\subset \mathcal{I}$ for all $\mu$ \cite{Baumgratz}.

With the definition of strictly incoherent operations, we then introduce the notion of stochastic strictly incoherent operations \cite{Liu5}. A stochastic strictly incoherent operation is constructed by a subset of strictly incoherent Kraus operators. Without loss of generality, we denote the subset as $\{K_{1},K_{2},\dots, K_{L}\}$. Otherwise, we may renumber the subscripts of these Kraus operators. Then, a stochastic strictly incoherent operation, denoted as $\Lambda_s(\rho)$, is defined by
\begin{equation}
\Lambda_s(\rho)=\frac{\sum_{\mu=1}^L K_\mu\rho K_\mu^{\dagger}}{\Tr(\sum_{\mu=1}^LK_\mu\rho K_\mu^{\dagger})},
\label{lams}
\end{equation}
where $\{K_{1},K_{2},\dots, K_{L}\}$ satisfies $\sum_{\mu=1}^L K_\mu^{\dagger}K_\mu\leq \mathbb{I}$. Clearly, the state $\Lambda_s(\rho)$ is obtained with probability $P=\Tr(\sum_{\mu=1}^LK_\mu\rho K_\mu^{\dagger})$ under a stochastic strictly incoherent operation $\Lambda_s$, while state $\Lambda(\rho)$ is fully deterministic under a strictly incoherent operation $\Lambda$. Here we emphasize that the stochastic transformation with $\sum_{\mu=1}^L K_\mu^{\dagger}K_\mu\leq I$ means that a copy of $\Lambda_s(\rho)$ may be obtained from a copy of $\rho$ with probability $\text{P}=\Tr(\sum_{\mu=1}^LK_\mu\rho K_\mu^{\dagger})(\leq1)$. That is, the stochastic transformation run the risk of failure with certain probability.

Since the notion of majorization is to play a central role in what follows, let us recall some notations about it \cite{Bhatia}. For the $d$-dimensional probability distributions $\mathcal{P}^d$, we say that a probability distribution $\textbf{p}=(p_1,p_2,...,p_d)$ is majorized by $\textbf{q}=(q_1,q_2,...,q_d)$, in symbols
\begin{eqnarray}
\textbf{p}\prec\textbf{q},
\end{eqnarray}
if  there are
\begin{eqnarray}
\sum_{i=1}^lp_i^\downarrow\leq\sum_{i=1}^lq_i^\downarrow,
\end{eqnarray}
for all $1\leq l\leq d-1$, where $\downarrow$  indicates that elements are to be taken in descending order.
Hereafter, we will apply majorization to density operators and write $\rho_1\prec\rho_2$ if the corresponding majorization relation holds for the eigenvalues of $\rho_1$ and $\rho_2$. More precisely, let $\rho_1$ and $\rho_2$ be two quantum states with their eigenvalues being $(\uplambda_1\geq\uplambda_2\geq\cdots\geq\uplambda_d)$ and $(\uplambda_1^\prime\geq\uplambda_2^\prime\geq\cdots\geq\uplambda_d^\prime)$, respectively. Then, $\rho_1\prec\rho_2$ implies that there are
\begin{eqnarray}
C_s(\rho_1):=\sum_{i=s}^d\uplambda_i\geq C_s(\rho_2):=\sum_{i=s}^d\uplambda_i^\prime
\end{eqnarray}
for all $1\leq s\leq d-1$ \cite{Bhatia}. Here and hereafter, we use $\rho^\downarrow$ to indicate that
the eigenvalues of $\rho$ are to be taken in descending order and we denote $\rho\prec\sum_np_n\rho_n^\downarrow$ as $C_s(\rho)\geq\sum_np_nC_s(\rho_n)$ for all $1\leq s\leq d-1$.

Finally, the fidelity of two states $\rho_1$ and $\rho_2$ is defined to be \cite{Jozsa}
\begin{eqnarray}
F(\rho_1,\rho_2)=\Tr\left(\sqrt{\sqrt{\rho_1}\rho_2\sqrt{\rho_1}}\right).
\end{eqnarray}
From Ref. \cite{Jozsa}, we have the following observations:\\
(1) $F(\ket{\varphi},\ket{\psi})=\abs{\langle\varphi\ket{\psi}}$;~~
(2) $F(\ket{\varphi},\rho)=\bra{\varphi}\rho\ket{\varphi}$.

\section{Approximation coherence distillation}

We start by presenting the basic task of the approximation coherence distillation via strictly incoherent operations:

For an initial state $\rho$, we want to transform it into a target state $\psi$. However, we may not achieve this task exactly. In this case, we may transform $\rho$ into some pure state $\phi$ or some pure state ensemble $\{p_\mu,\varphi_\mu\}$ whose fidelity with the target state $\psi$ is maximal instead.

For the sake of simplicity, we define $\mathcal{S}_{\rho}$ as the set of pure states $\phi$ or  pure state ensembles $\{p_\mu,\varphi_\mu\}$ which can be obtained from $\rho$ by using strictly incoherent operations and we denote $F_{\max}(\psi,\mathcal{S}_\rho)$ as the maximal fidelity achievable in the protocol.

First, we show that it is the pure coherent-state subspaces of $\rho$ that play the role in the approximate coherence distillation. To obtain this, let us recall the following result, which was given in Ref. \cite{Liu4}.

Theorem 1.--The transformation $\rho\to\{p_{\mu n},\ket{\varphi_{\mu n}}\}_n^\mu$ can be achieved by using strictly incoherent operations if and only if  there is an orthogonal and complete set of incoherent projectors $\{\mathbb{P}_\mu\}$ such that, for all $\mu$, there are
\begin{eqnarray}\label{condition}
\frac{\mathbb{P}_\mu\rho\mathbb{P}_\mu}{\Tr(\mathbb{P}_\mu\rho\mathbb{P}_\mu)}=\psi_\mu~\text{and}~\Delta\psi_\mu\prec\sum_np_{n|\mu}\Delta\varphi_{\mu n}^\downarrow,
\end{eqnarray}
where $\psi_\mu$ are pure states, $p_\mu=\Tr(\mathbb{P}_\mu\rho\mathbb{P}_\mu)$,  and $p_{n|\mu}:=p_{\mu n}/p_\mu$.

In particular, for $\rho$ being a pure state and some $p_{\mu n}=1$, we recover the results obtained in Ref. \cite{Chitambar2}.

Theorem $1^\prime$.--The transformation $\varphi\to\psi$ can be achieved by using strictly incoherent operations if and only if there is $\Delta\varphi\prec\Delta\psi$.

Thus from Theorem 1, we note that it is the parts of $\rho$ that $\mathbb{P}\rho\mathbb{P}$ being rank one are useful in the transformation $\rho\to\{p_{\mu},\ket{\varphi_{\mu}}\}_\mu$. For the sake of simplicity, we call these parts as the pure coherent-state subspaces of $\rho$. More precisely, if there is an incoherent projector $\mathbb{P}$ such that $\mathbb{P}\rho\mathbb{P}= \varphi$ with the coherence rank \cite{note} of $\varphi$ being $n\geq0$, then we say that $\rho$ has an $n+1$-dimensional pure coherent-state subspace corresponding to $\mathbb{P}$. Furthermore, we say that the pure coherent-state subspaces with the projector $\mathbb{P}$ for $\rho$ is maximal if the pure coherent-state subspace cannot be expanded to a larger one with an incoherent projector $\mathbb{P}^{\prime}$ such that $\mathbb{P}^{\prime}\rho\mathbb{P}^{\prime}= \varphi^{\prime}$, $\varphi^{\prime}\neq\varphi$, and $\mathbb{P}\varphi^{\prime}\mathbb{P}= \varphi$.

To identify the pure coherent-state subspaces of a state $\rho$, we resort to the following matrix
\begin{eqnarray}
\mathbb{A}=(\Delta\rho)^{-\frac12}\abs{\rho}(\Delta\rho)^{-\frac12}\label{A}.
\end{eqnarray}
Here, for the given state $\rho=\sum_{ij}\rho_{ij}\ket{i}\bra{j}$, the matrix $\abs{\rho}$ reads $|\rho|=\sum_{ij}|\rho_{ij}|\ket{i}\bra{j}$ and
$(\Delta\rho)^{-\frac12}$ is the diagonal matrix with elements
$(\Delta\rho)^{-\frac12}_{ii}= \left\{
  \begin{array}{ll}
     \rho_{ii}^{-\frac12}, &\text{if} ~ \rho_{ii}\neq0;\\
    0,&\text{if}~ \rho_{ii}= 0.
  \end{array}\right.$ A useful property of $\mathbb{A}$ to identify the pure coherent-state subspaces of $\rho$ can be presented as the following Theorem 2 (see Appendix I for details).

Theorem 2.--The rank of $\mathbb{P}\rho\mathbb{P}$ is $1$ if and only if all of its corresponding elements of $\mathbb{A}$ are $1$.

From this, we could obtain that if there are $n$-dimensional principal sub-matrices $\mathbb{A}_\mu$ of $\mathbb{A}$ with all its elements being $1$, then the corresponding subspace of $\rho$ is an $n$-dimensional pure coherent-state subspace. By using this result, one can easily identify the pure coherent-state subspaces of $\rho$. For a state $\rho$, let the corresponding Hilbert subspaces of principal submatrices $\mathbb{A}_\mu$ ($\mu=1,\cdots,\mathcal{U}$) be $\mathcal{H}_\mu$, which is spanned by
$\{\ket{i_1^\mu},\ket{i_2^\mu},\cdots,\ket{i_{d_\mu}^\mu}\}\subset\{\ket{1}, \ket{2},\cdots,\ket{d}\}$ and the corresponding incoherent projectors be $\mathbb{P}_\mu$, with its rank being $d_\mu$, i.e., \begin{eqnarray}
  \mathbb{P}_\mu=\ket{i^\mu_1}\bra{i^\mu_1}+\ket{i^\mu_2}\bra{i^\mu_2}+\cdots+\ket{i^\mu_{d_\mu}}\bra{i^\mu_{d_\mu}}.
\end{eqnarray}
Acting $\{\mathbb{P}_\mu\}$ on the state $\rho$, we then obtain the set $\{\varphi_\mu\}_{\mu=1}^\mathcal{U}$, where $\varphi_\mu$ have the form
$\varphi_\mu=(\mathbb{P}_\mu\rho\mathbb{P}_\mu)/\Tr(\mathbb{P}_\mu\rho\mathbb{P}_ \mu)$.
Let the pure states corresponding to maximal pure coherent-state subspaces be
\begin{eqnarray}
\frac{\mathbb{P}^\text{m}_\mu\rho\mathbb{P}^\text{m}_\mu}{\Tr(\mathbb{P}^\text{m}_\mu\rho\mathbb{P}^\text{m}_\mu)}=\varphi^\text{m}_\mu.
\end{eqnarray}
Here $\mathbb{P}^\text{m}_\mu$ are the incoherent projectors corresponding to maximal pure coherent-state subspaces.
Then, after acting the incoherent projectors $\{\mathbb{P}^\text{m}_\mu\}$ on $\rho$, we obtain a set of pure states $\varphi^\text{m}_\mu$ with probability $p_\mu=\Tr(\mathbb{P}^\text{m}_\mu\rho\mathbb{P}^\text{m}_\mu)$, i.e., there is
\begin{eqnarray}
\Lambda_\mathbb{P}(\rho)=\sum_{\mu=1}^\mathcal{U}\mathbb{P}^\text{m}_\mu\rho\mathbb{P}^\text{m}_\mu=\bigoplus_{\mu=1}^\mathcal{U} p_\mu\varphi_\mu^\text{m}. \label{mixed_first}
\end{eqnarray}
By Theorem 1 and the definitions of $\{\mathbb{P}^\text{m}_\mu\}$ and $\{\mathbb{P}_\mu\}$, it is apparent to see that, to obtain a pure state or a pure state ensemble from the state $\rho$, we only need to consider the state
\begin{eqnarray}
\rho^\text{m}=\bigoplus_{\mu=1}^\mathcal{U} p_\mu\varphi_\mu^\text{m}, \label{immediate}
\end{eqnarray}
since general $\rho^\prime=\bigoplus_{\mu}p_\mu\varphi_\mu$ can be obtained from $\rho^\text{m}$ by using strictly incoherent operations.

The results presented above imply that, to obtain some pure state $\phi$ or some pure state ensemble $\{p_\mu,\phi_\mu\}$ whose fidelity with the target state $\psi$ is maximal from $\rho$, we only need to study each $\varphi_\mu^\text{m}$.

Second, for an initial pure state $\varphi$ and a target pure state $\psi$, let us calculate the maximal fidelity achievable in the approximate coherence distillation.

For the pure coherent state $\varphi$, it may be transformed into a pure state $\phi$ or a pure state ensemble $\{p_\mu,\varphi_\mu\}$ by using strictly incoherent operations. We show that for the problem considered here, we only need to consider the former case. This leads to the following Theorem 3 (see Appendix II for details).

Theorem 3.--Let us define $\bar{F}\left(\psi,\{p_\mu,\varphi_\mu\}\right):=\sum_\mu p_\mu F(\psi,\varphi_\mu)$, where $\{p_\mu,\varphi_\mu\}$ is a pure state ensemble obtained from $\varphi$ by using strictly incoherent operations. Then, the maximum of $\bar{F}\left(\psi,\{p_\mu,\varphi_\mu\}\right)$ can always be obtained by $F(\psi,\phi)$ with $\phi\in\mathcal{S}_\varphi$.

With the above Theorem 3, for an initial pure state $\varphi$ and a target pure state $\psi$, to obtain the intermediate state achieving the maximal fidelity with $\psi$, we only need to consider the pure state in $\mathcal{S}_\varphi$. Next, we are ready to present the maximal fidelity and the intermediate state. To this end, we introduce some elementary notations. Let $\ket{\varphi}=\sum_{i=1}^d\varphi_i\ket{i}$ and $\ket{\psi}=\sum_{i=1}^d\psi_i\ket{i}$ be two pure coherent states with $\abs{\varphi_1}\geq \abs{\varphi_2}\geq\cdots\geq\abs{\varphi_d}$ and $\abs{\psi_1}\geq \abs{\psi_2}\geq\cdots\geq\abs{\psi_d}$, respectively. For the state $\ket{\varphi}$, we define $C_s^\varphi$ as $C_s^\varphi=\sum_{i=s}^d\abs{\varphi_i}^2$. Let us denote $s_1\in\{1,\cdots,d\}$ as the smallest integer such that
\begin{eqnarray}
q_1=\frac{C_{s_1}^\varphi}{C_{s_1}^\psi}:=
\min_s\frac{C_{s}^\varphi}{C_{s}^\psi}.
\end{eqnarray}
We should note that there may be the case that $q_1=1$ and $s_1=1$ at the same time. If this is not the case, for any $a,b,c,d\in\mathbb{R}^+$, the equivalence of $a(b+d)<b(a+c)$ and $ad<bc$ implies
that for any integer $s\in[1,s_1]$
\begin{eqnarray}
\frac{C_s^\varphi-C_{s_1}^\varphi}{C_s^\psi-C_{s_1}^\psi}>q_1.
\end{eqnarray}
Let us then denote $s_2\in[1,s_1-1]$ as the smallest integer such that
\begin{eqnarray}
q_2=\frac{C_{s_2}^\varphi-C_{s_1}^\varphi}{C_{s_2}^\psi-C_{s_1}^\psi}
:=\min_s\frac{C_s^\varphi-C_{s_1}^\varphi}{C_s^\psi-C_{s_1}^\psi},
\end{eqnarray}
where $q_2>q_1$. Repeating this process until $s_k=1$ for some $k$, we obtain a series of $k+1$ integers $s_0>s_1>s_2>\cdots>s_k$ ($s_0:=d+1$) and $k$ positive real numbers $0<q_1<q_2<\cdots<q_k$, by means of which we denote the final state as
\begin{eqnarray}
\ket{\phi}:=\sum_{i=1}^d\phi_i\ket{i}~~\text{with}~~\phi_i:=q_j\psi_i~~\text{if}~i\in[s_j,s_{j-1}-1]. \label{phi}
\end{eqnarray}
It is direct to examine that $\abs{\phi_i}\geq\abs{\phi_{i+1}}$ and there are
\begin{eqnarray}
C_s^\varphi\geq C_s^\phi,~~\forall s\in[1,d],
\end{eqnarray}
i.e., $\Delta\varphi\prec\Delta\phi$. By Theorem $1^\prime$, this means that the state $\ket{\varphi}$ can be transformed into $\ket{\phi}$ by using some strictly incoherent operation with certainty. Let us further define positive quantities
\begin{eqnarray}
A_j&&:=C_{s_j}^\varphi-C_{s_{j-1}}^\varphi=\sum_{i=s_j}^{s_{j-1}-1}\abs{\varphi_i}^2, \nonumber\\
B_j&&:=C_{s_j}^\psi-C_{s_{j-1}}^\psi=\sum_{i=s_j}^{s_{j-1}-1}\abs{\psi_i}^2,
\end{eqnarray}
where we have assumed that $C_{s_0}^\varphi=0$ and $C_{s_0}^\psi=0$. By using Eq. (\ref{phi}), we immediately derive the fidelity between $\ket{\phi}$ and the target state $\ket{\psi}$ is
\begin{eqnarray}
F(\phi,\psi)=\sum_{j=1}^k\sqrt{A_jB_j}. \label{optimal}
\end{eqnarray}

With the above notations, we arrive at Theorem 4.

Theorem 4.--For the initial state $\ket{\varphi}=\sum_{i=1}^d\varphi_i\ket{i}$ and the target state $\ket{\psi}=\sum_{i=1}^d\psi_i\ket{i}$ with $\abs{\varphi_1}\geq \abs{\varphi_2}\geq\cdots\geq\abs{\varphi_d}$ and $\abs{\psi_1}\geq \abs{\psi_2}\geq\cdots\geq\abs{\psi_d}$, respectively, there is
\begin{eqnarray}
F_{\max}(\psi,\mathcal{S}_\varphi):=\max_{\epsilon\in\mathcal{S}_\varphi}F(\psi,\epsilon)=\sum_{j=1}^k\sqrt{A_jB_j}.
\end{eqnarray}
The intermediate state achieving $F_{\max}(\psi,\mathcal{S}_\varphi)$ is the state $\ket{\phi}$ presented in Eq. (\ref{phi}).

\emph{Proof.}--Let $\ket{\epsilon}=\sum_{i=1}^d\epsilon_i\ket{i}$ with $\abs{\epsilon_i}\geq\abs{\epsilon_{i+1}}$ be an arbitrary pure state belonging to the set $\mathcal{S}_\varphi$. By direct calculations, we obtain $F(\epsilon,\psi)=\sum_{i=1}^d\abs{\epsilon_i\psi_i}$. By $\bm{a}\cdot \bm{b}\leq\abs{\bm{a}}\abs{\bm{b}}$, we derive
\begin{eqnarray}
F(\omega,\psi)=\sum_{i=1}^d\abs{\epsilon_i\psi_i}\leq
\sum_{j=1}^k\sqrt{A_j^\prime B_j},
\end{eqnarray}
where we have defined $A_j^\prime:=\sum_{i=s_j}^{s_{j-1}-1}\abs{\epsilon_i}^2$. Since $\epsilon$ can be obtained by strictly incoherent operations from $\varphi$, then there are $C_s^\epsilon\leq C_s^\varphi$ for all $s$.
We further define $x_j$ as
\begin{eqnarray}
x_j:=C_{s_j}^\varphi-C_{s_j}^{\epsilon}.
\end{eqnarray}
The condition $\epsilon\in\mathcal{S}_\varphi$ implies that $x_j\geq0$ for all $j$. Let us further define a function
\begin{eqnarray}
f(\textbf{x}):=\sum_{j=1}^k\sqrt{(A_j-x_j+x_{j-1})B_j}.
\end{eqnarray}
Next, we present that $f(\textbf{x})$ obtains its maximum at $\textbf{x}=\textbf{0}$ by showing that the Hessian is negative semidefinite at $\textbf{0}$ \cite{Horn}.
By direct calculations, we immediately derive
\begin{eqnarray}
\frac{\partial f(\textbf{x})}{\partial x_j}=\left(\sqrt{\frac{B_{j+1}}{A_{j+1}-x_{j+1}+x_j}}-\sqrt{\frac{B_{j}}{A_{j}-x_j+x_{j-1}}}\right).
\end{eqnarray}
We note that $\frac{\partial f(\textbf{0})}{\partial x_j}$ are negative for all $j$ since there are $\frac{A_j}{B_j}<\frac{A_{j+1}}{B_{j+1}}$. Further, we can derive that
the Hessian matrix $H:=[\frac{\partial^2f(\textbf{x})}{\partial x_i\partial x_j }]$ reads
\begin{eqnarray}\nonumber
H=\begin{pmatrix}
    -(z_1+z_2)&z_2&0&\cdots&0\\
    z_2&-(z_2+z_3)&z_3&\cdots&0\\
    0& z_3&-(z_3+z_4)&\cdots&0\\
    \vdots& \vdots& \vdots&\ddots&\vdots\\
    0&0&0&\cdots&-(z_k+z_{k+1})
  \end{pmatrix},
\end{eqnarray}
where $H_{jj}=-(z_j+z_{j+1})$, $H_{j(j-1)}=z_j$, and $H_{j(j+1)}=z_{j+1}$ with $z_j=\frac14\sqrt{B_j}(A_j-x_j+x_{j-1})^{-3/2}$. Finally, we show that $H$ is negative semidefinite. To this end, let us recall the Ger\v{s}gorin disk theorem, which says that if $H=[H_{ij}]$, then there is
\begin{eqnarray}
\{\uplambda(H)\}\subset G(H)=\bigcup_{n=1}^NG_n(H),
 \end{eqnarray}
 where $\uplambda(H)$ are the eigenvalues of $H$ and
 \begin{eqnarray}
 G_n(H):=\{z\in\mathbb{C}:\abs{z-H_{nn}}\leq\sum_{j\neq n}\abs{H_{nj}}\}.
 \end{eqnarray}
  From this, we immediately derive
\begin{eqnarray}
\abs{\uplambda-H_{nn}}\leq\abs{H_{nn-1}}+\abs{H_{nn+1}}.
 \end{eqnarray}
 Then, for all $i$, there are
 \begin{eqnarray}
 -(z_j+z_{j+1})\leq\uplambda(H)\leq0.
  \end{eqnarray}
Thus the matrix $H$ is negative semidefinite. This further implies that $f(\textbf{x})$ obtains the maximum at $\textbf{x}=\textbf{0}$.
Henceforth, we obtain
\begin{eqnarray}
f(\textbf{x})_{\max}=\sum_{j=1}^k\sqrt{A_jB_j}.
\end{eqnarray}
Since the state $\phi$ defined in Eq. (\ref{phi}) could achieve this maximum, we complete the proof of this theorem.~~~~~~~~~~~~~~~~~~~~~~~~$\blacksquare$

Finally, we present the approximate coherence distillation for a mixed state.

From Theorem 1, for a state $\rho$, we only need to consider the protocol of the state $\rho^\text{m}=\bigoplus_{\mu=1}^\mathcal{U} p_\mu\varphi_\mu^\text{m}$ in Eq. (\ref{immediate}). Thus for the initial state $\rho$ and the target state $\psi$, we should calculate each $F_{\max}(\psi, \mathcal{S}_{\varphi_\mu^\text{m}})$ as in Theorem 3, respectively. Then, we immediately obtain
\begin{eqnarray}
F_{\max}(\psi,\mathcal{S}_\rho)=\min_\mu F_{\max}(\psi, \mathcal{S}_{\varphi_\mu^\text{m}}).
\end{eqnarray}

We then summarize the above results as Theorem 5.

Theorem 5.--For an initial state $\rho$ and a target state $\psi$, let the state corresponding to its maximal pure coherent-state subspaces be $\Lambda_\mathbb{P}(\rho)=\bigoplus_{\mu=1}^\mathcal{U} p_\mu\varphi_\mu^{\text{m}}$. The maximal fidelity achievable by using strictly incoherent operations is
\begin{eqnarray}
F_{\max}(\psi,\mathcal{S}_\rho)=\min_\mu F_{\max}(\psi, \mathcal{S}_{\varphi_\mu^\text{m}}),
\end{eqnarray}
where  $\mathcal{S}_{\rho}$ is the set of pure states that can be obtained from $\rho$ by using strictly incoherent operations.

In particular, it is reminiscent of the case of entanglement \cite{Vidal}, where the approximate transformations of pure entangled states were studied with the fidelity being $F(\rho_1,\rho_2)=\Tr\left(\sqrt{\sqrt{\rho_1}\rho_2\sqrt{\rho_1}}\right)^2$.

We point out that, for pure states, the results presented in Theorem 4 can be naturally extended to incoherent operations by following the same arguments around Theorem 4. However, this is not the case for mixed states. To see this, let us consider the initial state $\rho$ as
\begin{eqnarray}\label{state}
  \rho=\begin{pmatrix}
    \frac14&0&\frac1{2\sqrt{5}}&\frac1{4\sqrt{5}}\\
    0&\frac14&-\frac1{4\sqrt{5}}&\frac1{2\sqrt{5}}\\
   \frac1{2\sqrt{5}}&-\frac1{4\sqrt{5}}&\frac14&0\\
    \frac1{4\sqrt{5}}&\frac1{2\sqrt{5}}& 0&\frac14
  \end{pmatrix},
\end{eqnarray}
and the target state as $\ket{\psi}=\frac1{\sqrt{2}}(\ket{1}+\ket{2})$. Then, by direct calculations, $F_{\max}(\psi,\mathcal{S}_\rho)=\frac1{\sqrt{2}}$. However, by using the incoherent operations $\Lambda(\cdot)=K_1(\cdot)K_1^\dag+K_2(\cdot)K_2^\dag$ with
\begin{eqnarray}
K_1=\begin{pmatrix}
    \frac45&\frac35&0&0\\
    0&0&\frac1{\sqrt{5}}&\frac2{\sqrt{5}}\\
    0& 0&0&0\\
    0& 0& 0&0
  \end{pmatrix},~~
 K_2=\begin{pmatrix}
    -\frac35&\frac45&0&0\\
    0&0&-\frac2{\sqrt{5}}&\frac1{\sqrt{5}}\\
    0& 0&0&0\\
    0& 0& 0&0
  \end{pmatrix},
\end{eqnarray}
 the maximal fidelity achievable is  $F_{\max}=1$.

\section{Remarks and conclusions}

Before concluding, we would like to compare the approximation coherence distillation with the deterministic coherence distillation \cite{Liu1} and the probabilistic coherence distillation \cite{Liu2}. To this end, let us consider the following question: (i) For an initial state $\rho$, if we want to transform it into a some pure state $\phi$ whose fidelity with the target state $\psi$ is equal to or larger than some value $F_0$, then can we achieve this task?  By using Theorem 5, we only need to compare if there is
\begin{eqnarray}
F_{\max}(\psi,\mathcal{S}_\rho)\geq F_0.
\end{eqnarray}
Suppose $F_{\max}(\psi,\mathcal{S}_\rho)\geq F_0$, then we can achieve the task successfully. Conversely, if $F_{\max}(\psi,\mathcal{S}_\rho)< F_0$, then we cannot achieve this task with certainty. In the latter case, we could consider the problem (ii): If we cannot complete this distillation scheme with certainty, then what is the maximal probability of success in such a transformation? For this problem, let us define the set $S:=\{\mu|F_{\max}(\psi, \mathcal{S}_{\varphi_\mu^\text{m}})\geq F_0\}$. Then, the maximal probability of success in such a transformation is
\begin{eqnarray}
\text{P}_{\max}=\sum_{\mu\in S}p_\mu.
\end{eqnarray}
In particular, if the desired fidelity $F_0$ is $1$, then we can recover the results of deterministic coherence distillation \cite{Liu1} and the probabilistic coherence distillation \cite{Liu2}.

We should note that the intermediate state $\phi$ may not be a coherent state. Thus  given the initial state $\rho$, the target state $\psi$, and the desired fidelity $F_0$, we can decide whether the protocol is useful or not by comparing $F_{\max}(\psi,\mathcal{S}_\rho)$ with $F_0$.

To summarize, we have characterized the framework of approximate coherence distillation for a general state $\rho$. The aim of this protocol is to obtain an intermediate state by using strictly incoherent operations from $\rho$ most approximate to the target state $\psi$ in terms of fidelity. We have presented the explicit conversion strategy of the approximate coherence distillation. This distillation protocol can be seen as a generalization of the exact coherence distillation.

In passing, we would like to point out that the situation we consider here is different from the one-shot coherence distillation developed in Refs. \cite{Fang,Regula1}, where the intermediate state is not necessarily a pure state.

\section*{Acknowlegements}

This work is supported by the NSFC (Grant No. 12088101), and NSAF (Grant No. U1930403, No. U1930402). C.L.L acknowledges support from the China Postdoctoral Science Foundation Grant No. 2021M690324.

\section*{Appendix I}

Theorem 2. Let $\mathbb{A}:=(\Delta\rho)^{-\frac12}\abs{\rho}(\Delta\rho)^{-\frac12}$. Then $\rho$ is a pure state if and only if all the elements of $\mathbb{A}$ are $1$.

\emph{Proof.}--Let $\rho=\ket{\phi}\bra{\phi}$ with $\ket{\phi}=\sum_{i=1}^d\phi_i\ket{i}$ being a pure state. Then there are $\rho_{ij}=\phi_i\phi_j^*$, $\rho_{ii}=\abs{\phi_i}^2$, and $\rho_{jj}=\abs{\phi_j}^2$. It is direct to examine that all the elements of $\mathbb{A}$ are $1$, i.e., \begin{eqnarray}
\mathbb{A}_{ij}=\sqrt{\frac{\abs{\rho_{ij}}^2}{\rho_{ii}\rho_{jj}}}=1
\end{eqnarray}
for all $i,j$. This completes the \emph{only~if} part of the Theorem.

For the \emph{if} part of the Theorem, let us consider the matrix
\begin{eqnarray}
\mathbb{A}^\prime:=(\Delta\rho)^{-\frac12}\rho(\Delta\rho)^{-\frac12}
\end{eqnarray}
 with $\abs{\mathbb{A}^\prime}=\mathbb{A}$. It is direct to see that the matrix $\mathbb{A}^\prime$ is positive semidefinite and thus $\mathbb{A}^\prime_{ii}=\mathbb{A}_{ii}=1$ for all $i$. Since all the elements of $\mathbb{A}$ are 1, then there are $\abs{\mathbb{A}^\prime_{ij}}=1$ for all $i,j$. Without loss of generality, let us assume that $\mathbb{A}=\sum_\mu\ket{\varphi_\mu}\bra{\varphi_\mu}$ with
$\ket{\varphi_\mu}=\sum_i\varphi_i^\mu\ket{i}$. Then,
by direct calculations, we obtain
\begin{eqnarray}
\abs{\mathbb{A}_{ij}^\prime}^2=\abs{\sum_\mu{\varphi^\mu_i}^*\varphi_j^\mu}^2
\leq\sum_\mu\abs{\varphi^\mu_i}^2\sum_\mu\abs{\varphi^\mu_j}^2=\mathbb{A}_{ii}^2\mathbb{A}_{jj}^2,
\end{eqnarray}
with equality if and only if, for some $k_{ij}\in\mathbb{C}$, there are \begin{eqnarray}
(\varphi^1_i,\varphi^2_i,\cdots,\varphi^d_i)=k_{ij}(\varphi^1_j,\varphi^2_j,\cdots,\varphi^d_j).
\end{eqnarray}
 This implies that there are $\ket{\varphi_\mu}\bra{\varphi_\mu}=\abs{k_{\mu\nu}}^2\ket{\varphi_\nu}\bra{\varphi_\nu}$. Thus, the rank of $\mathbb{A^\prime}$ is 1. This further implies that $\rho$ is rank 1. To see this, let us show that if A, B are Hermitian operators, if A is invertible on $\mathbb{M}_n$, then there is
\begin{eqnarray}
\text{Rank}(B)=\text{Rank}(ABA^\dag).
\end{eqnarray}
On the one hand, let $B=\sum_{i=1}^\Lambda\uplambda_i\ket{\uplambda_i}\bra{\uplambda_i}$. Then there is $ABA^\dag
=\sum_{i=1}^\Lambda\uplambda_i\ket{\psi_i}\bra{\psi_i},$ where $\ket{\psi_i}:=A\ket{\uplambda_i}$. This means that
\begin{eqnarray}
\text{Rank}(B)\geq\text{Rank}(ABA^\dag).
\end{eqnarray}
On the other hand, since $A$ is invertible, let $C:=A^{-1}$. Then, there is $B=CABA^{\dag}C^\dag=\sum_{i=1}^\Lambda\uplambda_i\ket{\uplambda_i}\bra{\uplambda_i}.$
Thus, we obtain
\begin{eqnarray}
\text{Rank}(B)\leq\text{Rank}(ABA^\dag).
 \end{eqnarray}
 From this, since the rank of $\mathbb{A^\prime}$ is 1, then $\rho$ is also of rank 1. i.e., $\rho$ is a pure state. This completes the proof of the \emph{if} part of the Theorem.~~~~~~~~~~~~~~~~~~~~~~~~~~~~~~~~~~~~~~~~~~~~~~~~~~~~~~~~~~~~~~~~~~~~~~~~~~~~~~~~~~~$\blacksquare$

\section*{Appendix II}

Theorem 3.--Let us define $\bar{F}\left(\psi,\{p_\mu,\varphi_\mu\}\right):=\sum_\mu p_\mu F(\psi,\varphi_\mu)$, where $\{p_\mu,\varphi_\mu\}$ is an pure state ensemble obtained from $\varphi$ by using strictly incoherent operations. Then, the maximum of $\bar{F}\left(\psi,\{p_\mu,\varphi_\mu\}\right)$ can always be obtained by $F(\psi,\phi)$ with $\phi\in\mathcal{S}_\varphi$.

\emph{Proof.}--We first show that for two pure states $\ket{\varphi}=\sum_{i=1}^d\varphi_i\ket{i}$ and $\ket{\psi}=\sum_{i=1}^d\psi_i\ket{i}$, there are
\begin{eqnarray}
F_m^\mathbb{U}:=\max_{\mathbb{U}}F(\varphi,\mathbb{U}\psi\mathbb{U}^\dag)=
\sum_{i=1}^d\abs{\varphi_i^\downarrow}\abs{\psi_i^\downarrow},
\end{eqnarray}
where the maximum is taken over incoherent unitary operations $\mathbb{U}$.

To this end, an incoherent unitary $\mathbb{U}$ can be expressed as
\begin{eqnarray}
\mathbb{U}=\sum_{j=1}^de^{i\theta_j}\ket{\pi(j)}\bra{j},
\end{eqnarray}
where $\pi$ is a permutation of $\{1,\cdots,d\}$. Then, we derive
\begin{eqnarray}
F(\varphi,\mathbb{U}\psi\mathbb{U}^\dag)=\abs{\bra{\varphi}\mathbb{U}\ket{\psi}}=
\abs{\sum_{j=1}^de^{i\theta_{\pi(j)}}\varphi_{\pi(j)}\psi_j}.
\end{eqnarray}
By using the triangle inequality, we obtain
\begin{eqnarray}
\abs{\sum_{j=1}^de^{i\theta_{\pi(j)}}\varphi_{\pi(j)}\psi_j}\leq\sum_j\abs{\varphi_{\pi(j)}\psi_j}.
\end{eqnarray}
By using a result in Ref. \cite{Bhatia} which says that, for any two $d$-dimensional real vectors $\bm{a},\bm{b}$, there is $\langle \bm{a},\bm{b} \rangle\leq\langle \bm{a}^\downarrow,\bm{b}^\downarrow \rangle$ where $\downarrow$ indicates that the elements are to be taken in descending order, we get
\begin{eqnarray}
\sum_j\abs{\varphi_{\pi(j)}\psi_j}\leq\sum_{i=1}^d\abs{\varphi_i^\downarrow}\abs{\psi_i^\downarrow},
\end{eqnarray}
where the equality can be achieved by choosing the incoherent unitary such that $\abs{\varphi_1}\geq \abs{\varphi_2}\geq\cdots\geq\abs{\varphi_d}$ and $\abs{\psi_1}\geq \abs{\psi_2}\geq\cdots\geq\abs{\psi_d}$ and $\abs{\varphi_j^\downarrow}\abs{\psi_j^\downarrow}=e^{i\theta_j}\varphi_j\psi_j$.

From the results presented above, without loss of generality, we assume that $\ket{\varphi_\mu}=\sum_{i=1}^{d_\mu}c_i^\mu\ket{i}$ with $\abs{c_{i}^\mu}\geq\abs{c_{i+1}^\mu}\geq0$ and $\ket{\psi}=\sum_{i=1}^d\psi_i\ket{i}$ with $\abs{\psi_i}\geq\abs{\psi_{i+1}}\geq0$. Since $\{p_\mu,\varphi_\mu\}$ is obtained from $\varphi$ by using strictly incoherent operations, then there are \cite{Chitambar2}
\begin{eqnarray}
\sum_\mu p_\mu \sum_{i=l}^{d_\mu}\abs{c_i^\mu}^2\leq\sum_{i=l}^{d_\mu}\abs{\psi_i}^2. \label{trans}
\end{eqnarray}
By direct calculations, we obtain
\begin{eqnarray}
\bar{F}_\text{max} \left(\psi,\{p_\mu,\varphi_\mu\}\right)&&=\sum_np_n\abs{\sum_i\psi_ic_{i}^\mu}\nonumber\\&&\leq \sum_\mu p_\mu\sum_{i=1}^n\abs{c_{i}^\mu\psi_i}, \label{fide_1}
\end{eqnarray}
where the maximum is taken over all ensemble $\{p_\mu,\varphi_\mu\}$ that can be obtained from $\varphi$ by using strictly incoherent operations and we have used the triangle inequality for the inequality.
On the other hand, let us consider the state
\begin{eqnarray}
\ket{\phi}=\sum_{i=1}^n\sqrt{\sum_\mu p_\mu\abs{c_{i}^\mu}^2}\ket{i}.
\end{eqnarray}
By using the relations in Eq. (\ref{trans}) and Theorem $1^\prime$, it is direct to examine that the transformation from $\varphi$ into $\phi$ can be realized with certainty by using strictly incoherent operations. Thus, the fidelity of $\ket{\psi}$ and $\ket{\phi}$ is
\begin{eqnarray}
F(\psi,\phi)=\sum_{i=1}^n\abs{\psi_i\sqrt{\sum_\mu p_\mu\abs{c_{i}^\mu}^2}}.
\end{eqnarray}
Applying the concavity of the function $f(x)=\sqrt{x}$, we obtain
\begin{eqnarray}
\sum_{i=1}^n\abs{\psi_i}\abs{\sqrt{\sum_\mu p_\mu\abs{c_{i}^\mu}^2}}\geq\sum_{i=1}^n\abs{\psi_i}\sum_\mu p_\mu\abs{c_{i}^\mu}.\label{fide_2}
\end{eqnarray}
From Eqs. (\ref{fide_1}) and (\ref{fide_2}), we immediately obtain that
\begin{eqnarray}
F(\psi,\phi)\geq\bar{F}\left(\psi,\{p_\mu,\varphi_\mu\}\right).
\end{eqnarray}
This completes the proof of the Theorem. ~~~~~~~~~~~~~~~~~~~~~~~~~~~~~~~~~~~~~~$\blacksquare$


\begin{thebibliography}{99}
\bibitem{note}For a pure state $\ket{\varphi}=\sum_{i=1}^Rc_i\ket{i}$ with $c_i\neq 0$, the coherence rank \cite{Liu5} $C_R$ of it is defined as the number of nonzero terms in this decomposition minus $1$, i.e., $C_R(\varphi)=R-1$.
\end{thebibliography}
\end{document}